\documentclass[preprint,12pt,3p]{elsarticle}

\usepackage[british]{babel}

\usepackage{hyphenat}

\usepackage{booktabs}

\usepackage{gensymb}

\usepackage{siunitx}

\usepackage{color,soul}

\usepackage[dvipsnames]{xcolor}

\usepackage[version=4]{mhchem}

\usepackage{subfig}

\usepackage[figuresright]{rotating}

\usepackage{multirow}

\usepackage{setspace}

\usepackage{hyperref}
\hypersetup{
	colorlinks   = true, 
	urlcolor     = darkorange, 
	linkcolor    = RoyalBlue, 
	citecolor    = RoyalBlue 
}

\usepackage{adjustbox}

\biboptions{sort&compress}

\usepackage[sfdefault,condensed]{roboto}

\usepackage[italic]{mathastext}

\usepackage[T1]{fontenc}

\begin{document}
	
	\begin{frontmatter}
		
		\journal{arXiv}
		
		\title{\raggedright\Large\textbf{Enhancing Irradiation Resistance in Refractory Medium Entropy Alloys with Simplified Chemistry}}
		
		\author[MUL]{M.A. Tunes\corref{cor}}\ead{matheus.tunes@unileoben.ac.at}
		\cortext[cor]{Corresponding authors:}	
		\author[LANL,UCB]{D. Parkison} 
            \author[CLEM]{B. Sun}
		\author[MUL]{P. Willenshofer}
		\author[MUL]{S. Samberger}
		\author[CINT]{B.K. Derby}
		\author[CINT]{J.K.S. Baldwin}
        \author[CINT]{S.J. Fensin}
		\author[WUT]{D. Sobieraj}
		\author[WUT]{J.S. Wr\'{o}bel}
        \author[UH]{J. Byggm\"astar}
		\author[MUL]{S. Pogatscher}
        \author[CLEM]{E. Martinez}
		\author[UKAEA,OXMAT]{D. Nguyen-Manh}
		\author[PNNL]{O. El-Atwani}

  
		\address[MUL]{Chair of Non-Ferrous Metallurgy, Montanuniversit\"at Leoben, Austria}
		\address[LANL]{Materials Science and Technology Division, Los Alamos National Laboratory, United States}
  		\address[UCB]{Department of Nuclear Engineering, University of California at Berkeley, United States }
    	\address[CLEM]{Departments of Mechanical Engineering and Materials Science and Engineering, Clemson University, United States}
		\address[CINT]{Center for Integrated Nanotechnologies, Los Alamos National Laboratory, United States}
        \address[WUT]{Faculty of Materials Science and Engineering, Warsaw University of Technology, Poland}
        \address[UH]{Department of Physics, University of Helsinki, Finland}
        \address[UKAEA]{Materials Division, United Kingdom Atomic Energy Authority, Culham Campus, Abingdon, United Kingdom}
        \address[OXMAT]{Department of Materials, University of Oxford, Parks Road, United Kingdom}
		\address[PNNL]{Reactor Materials and Mechanical Design, Pacific Northwest National Laboratory, United States}
		
		\begin{abstract}
        \singlespacing
			\noindent Refractory High-Entropy Alloys (RHEAs) hold promising potential to be used as structural materials in future nuclear fusion reactors, where W and its alloys are currently leading candidates. Fusion materials must be able to withstand extreme conditions \cite{zinkle2005fusion,zinkle2013challenges}, such as (i) severe radiation-damage arising from highly-energetic neutrons, (ii) embrittlement caused by implantation of H and He ions, and (iii) exposure to extreme high-temperatures and thermal gradients. Recent research demonstrated that two RHEAs -- the WTaCrV \cite{el2019outstanding} and WTaCrVHf \cite{el2023quinary} -- can outperform both coarse-grained and nanocrystalline W in terms of its radiation response and microstructural stability \cite{el2014situ,el2014ultrafine,el2017role,el2018nanohardness}. Chemical complexity and nanocrystallinity enhance the radiation tolerance of these new RHEAs, but their multi-element nature, including low-melting Cr, complicates bulk fabrication and limits practical applications. Herein, we demonstrate that reducing the number of alloying elements and yet retain high-radiation tolerance is possible within the ternary system W--Ta--V via synthesis of two novel nanocrystalline refractory medium-entropy alloys (RMEAs): the W$_{53}$Ta$_{44}$V$_{3}$ and W$_{53}$Ta$_{42}$V$_{5}$ (in at.\%). We experimentally show that the radiation response of the W--Ta--V system can be tailored by small additions of V, and such experimental result was validated with theoretical analysis of chemical short-range orders (CSRO) from combined ab-initio atomistic Monte-Carlo modeling. It is predicted from computational analysis that a small change in V concentration has a significant effect on the Ta-V CRSO between W$_{53}$Ta$_{44}$V$_{3}$ and W$_{53}$Ta$_{42}$V$_{5}$ leading to radiation-resistant microstructures in these RMEAs from chemistry stand-point of views. By returning to terminal solid solutions, we deviate from the original high-entropy alloy concept to show that high radiation resistance can be achieved in systems with simplified chemical complexity, hinting at a new paradigm for the metallurgy of high-entropy and highly-concentrated multi-component alloys.
      
		\end{abstract}
		
		\begin{keyword}
			Refractory Medium Entropy Alloys \sep Fusion Materials \sep Radiation Damage \sep High Entropy Alloys \sep \textit{In situ} TEM
		\end{keyword}
	\end{frontmatter}
	
\onehalfspacing
	
\section{Introduction}
\label{sec:introduction}
	
\noindent Since the 1970s, materials selection for application in future thermonuclear fusion reactors has posed a significant challenge for materials science and metallurgy \cite{bloom2004materials,zinkle2005advanced,duffy2010fusion,zinkle2013challenges,knaster2016materials,rowcliffe2018materials}. The deuterium-tritium fusion reaction exposes fusion reactors and their structural materials to exceptionally harsh environmental conditions. To date, considering the design and selection of materials for experimental fusion reactors, the following key factors must be addressed, not exhaustively \cite{bloom2004materials,zinkle2005advanced,duffy2010fusion,zinkle2013challenges,knaster2016materials,rowcliffe2018materials}:
	
	\begin{itemize}
		\item Impact of highly mono-energetic fusion neutrons (14.1 MeV) resulting in radiation damage;
		\item Implantation of helium (He) and hydrogen (H) ions at moderate-to-low energies after thermalization in the plasma, resulting in synergistic radiation damage, inert gas bubbles nucleation and growth and materials embrittlement; and
		\item The presence of a plasma with high-power density, resulting in high-temperature exposure in thermal gradients.
	\end{itemize}
	
As the 21st century began, two critical objectives arose that must be addressed to render fusion reactors feasible: (i) better controlling the high-power plasma with (ii) concomitant enhancement of its performance \cite{zinkle2013challenges}. For these reasons, materials science has redirected its attention towards better understanding plasma-materials interactions \cite{wirth2011fusion}. The surface of plasma-facing materials will be constantly subjected to the severe degradation forces aforementioned, which can significantly hinder the fusion reactor operation \cite{zinkle2013challenges}. Currently, beryllium (Be) \cite{zinkle2013challenges,sharp2024investigation} and tungsten (W) \cite{zinkle2013challenges}  are being considered for use in the first-wall armor, while W is the choice for divertor applications \cite{davis1998assessment,neu2005tungsten,rieth2011tungsten,wurster2013recent,rieth2013recent,ren2018methods}.
	
A significant challenge associated with the potential use of W in fusion reactors was highlighted in a 1998 article by Davis \textit{et al}. \cite{davis1998assessment}, which assessed W for application in the International Thermonuclear Experimental Reactor (ITER). The conclusion of this assessment stated that W was not suitable for use in ITER due to its low strength, limited thermal shock resistance, and high ductile-brittle transition temperature (DBTT) \cite{davis1998assessment}. Moreover, beyond the issues of low strength and brittleness, studies on energetic particle irradiation response have revealed a series of irradiation effects and microstructural changes in W that cannot be ignored \cite{harrison2017engineering,harrison2018damage,harrison2019use}. These irradiation effects, when emulated with particle accelerators -- as recently reviewed and summarized by Harrison \cite{harrison2019use} -- include a high density of dislocation loops and the formation of He bubbles \cite{yi2017study,yi2018high}, both of which can further increase W brittleness through irradiation-induced hardening. Ferroni \textit{et al.} also reported that when radiation damage is formed in W, even high temperature isochronal annealing is not able to fully recover defects in the W microstructure \cite{ferroni2015high}.
	
The challenge associated with improving mechanical properties of W has attracted increased attention of the greater materials and metallurgical engineering communities in the recent years \cite{ren2018methods,reiser2017ductilisation}. Various approaches have been explored, including reducing the grain size of W through methods such as cold-rolling \cite{reiser2017ductilisation,bonnekoh2018brittle,bonnekoh2019brittle,bonnekoh2020brittle}, wire drawing \cite{riesch2016properties,riesch2017tensile,nikolic2018effect,nikolic2018effecttwo}, and severe-plastic deformation (SPD) \cite{faleschini2005fracture,faleschini2007fracture}. Another avenue involves grain-boundary doping \cite{raabe2014grain,ren2018methods,wurmshuber2023enhancing,wurmshuber2023small}, although challenges associated with intragranular fracture and fracture resistance of nanocrystalline W are still pending resolution and can be considered topics for further research \cite{pippan2016importance,hohenwarter2015fracture,pineau2016failure,pineau2016failurethree}. Nevertheless, it is worth noting that recent radiation damage studies -- carried out with light- and heavy-ion irradiation within \textit{in situ} transmission electron microscopy (TEM) independently at both the IVEM facility in USA \cite{li2015tem} and the MIAMI facility in UK \cite{hinks2011miami,greaves2019new} -- have shown that even nanocrystalline W experience severe damage from energetic particle irradiation \cite{el2014situ,el2014ultrafine,el2017role,el2018nanohardness}, thus raising questions regarding the overall feasibility of proposing W for fusion applications. It is important emphasizing that \textit{in situ} TEM ion irradiation studies on coarse-grained W and select W-alloys also indicated extensive formation of radiation damage defects in a similar manner \cite{yi2013situ,yi2015characterisation,yi2016situ}. Neither an irradiation nor thermal stability study have yet been conducted on doped nanocrystalline W, indicating potential for future research still considering this element a candidate as fusion material.
	
An alternative to W in fusion applications is presented by the field of high-entropy alloys (HEAs) \cite{cantor2002novel,ranganathan2003alloyed,yeh2004nanostructured,cantor2004microstructural,miracle2017critical}, and more specifically due to their high-temperature resilience, the refractory high-entropy alloys (RHEAs) \cite{senkov2010refractory,gao2015design,senkov2018development,srikanth2021review}. Composed of four or more alloying elements in near-equimolar concentrations, these alloys are designed to maximize configurational entropy and minimize Gibbs free energy, thereby enhancing thermodynamic stability of the solid solution phase. Nanocrystalline RHEAs can be fabricated by several methods \cite{koch2017nanocrystalline}, including both SPD \cite{schuh2015mechanical,straumal2021severe} and mechanical alloying \cite{vaidya2019high} for macro-scale samples, and magnetron-sputtering \cite{braeckman2015high,tunes2018synthesis} for nano-scale prototypic samples: the latter aiming at fast irradiation screening. It is important to emphasize that not all nanocrystalline HEAs and RHEAs can be considered radiation-tolerant for fusion applications \cite{zhang2019thermal,tunes2021irradiation,tunes2023perspectives} and detailed studies should be carried out for each specific alloy under consideration.
	
Recently, two particular RHEAs systems have attracted the attention of the fusion materials community: the W--Ta--Cr--V \cite{el2019outstanding,tunes2023perspectives} and the W--Ta--Cr--V--Hf \cite{el2023quinary,tunes2023perspectives}. In the first quaternary system, the W$_{38}$Ta$_{36}$Cr$_{15}$V$_{11}$ RHEA (in at.\%), demonstrated superior radiation resistance compared to nanocrystalline W \cite{el2014situ,el2014ultrafine,el2017role,el2018nanohardness}, particularly in mitigating displacement damage formation such as dislocation loops, yielding negligible irradiation hardening assessed via nanomechanical testing. On the downside, this W$_{38}$Ta$_{36}$Cr$_{15}$V$_{11}$ RHEA experienced radiation-induced precipitation (RIP), characterized by the formation of Cr-V rich precipitates at a radiation dose of 8 dpa (displacement-per-atoms). In the second system, irradiation testing on the nanocrystalline RHEA -- the W$_{29}$Ta$_{42}$Cr$_{5}$V$_{16}$Hf$_{8}$ (in at.\%) revealed that neither dislocation loops nor precipitates have formed after a dose of 10 dpa \cite{el2023quinary}. Although such recent studies on these two RHEAs show that increased chemical complexity enhances radiation resistance and stability in harsh environments \cite{el2019outstanding,el2023quinary,tunes2023perspectives}, increasing the number of alloying elements in any RHEA system complicates their bulk fabrication, crucial for practical engineering applications: for example, the low-melting point of Cr impairs the feasibility for the fabrication of both W$_{38}$Ta$_{36}$Cr$_{15}$V$_{11}$ and W$_{29}$Ta$_{42}$Cr$_{5}$V$_{16}$Hf$_{8}$ RHEAs. Therefore, a key question arises: can the number of alloying elements be reduced in these RHEA systems without compromising their high radiation resistance?
	
In this study, we show that it is feasible to simplify the quinary RHEA system, the W--Ta--Cr--V--Hf, to a ternary refractory medium-entropy alloy (RMEA) system, the W--Ta--V, resulting in the development of a novel nanocrystalline RMEA, the W$_{53}$Ta$_{42}$V$_{5}$ (in at.\%).  A recent computational study indicated the WTaV system as a potential and most-promising low-activation RMEA with superior radiation resistance when compared with recent RHEAs proposed for application in irradiation environments \cite{wei2024revealing}. We demonstrate that the new ternary RMEA not only retains the irradiation resistance observed previously for more complex RHEAs \cite{el2019outstanding,tunes2023perspectives,el2023quinary}, but also surpasses the irradiation resistance of the binary ultrafine-grain (UFG) W$_{44}$Ta$_{56}$ (in at.\%) highly-concentrated refractory alloy. Through detailed post-irradiation analysis and atomistic Monte-Carlo (MC) simulations, we investigate the underlying mechanisms of irradiation resistance in this new ternary system, focusing on the role of minor V additions to the irradiation resistance output. We show both theoretically and experimentally that small additions of V drastically changes the non-irradiation resistant binary WTa system by forming a new irradiation resistant alloy -- the nanocrystalline ternary WTaV RHEA. In addition, we show demonstrate for the first time that the element Cr is not needed for RHEAs in the context of fusion applications, opening an unprecedented pathway for the synthesis of these novel metallic alloys in bulk and large-scale quantities.
	
\section{Results and Discussion}
\label{sec:discussion}
	
\subsection{Morphological stability under extreme conditions}
\label{sec:discussion:damage}
	
\noindent Heavy ion irradiation with \textit{in situ} TEM allowed for the comparison of two different RMEAs within the ternary system W--Ta--V, the W$_{53}$Ta$_{42}$V$_{5}$ and the W$_{53}$Ta$_{44}$V$_{3}$. It is important emphasizing that the major objective of the study was to identify the role of the element V in these two alloys' response to both high-temperature annealing (maximum temperature was 1173 K) and high-dose irradiation at high-temperatures (maximum average dose was 20 dpa at an irradiation temperature of 1073 K).
	
\begin{figure}[hb!]
		\centering	
		\includegraphics[width=\textwidth]{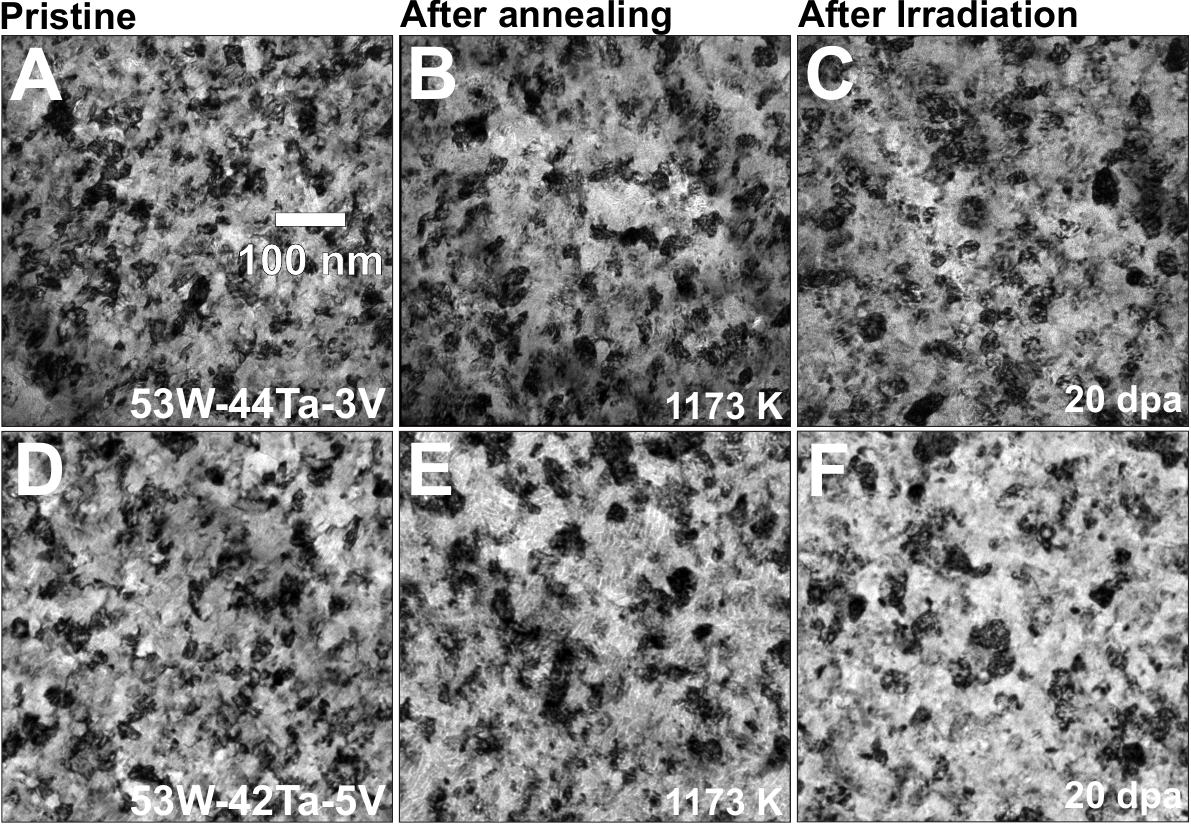}
		\caption{\textbf{Microstructural response to annealing and irradiation} | BFTEM micrographs taken at pristine condition, after annealing (at 1173 K) and after irradiation (at 1073 K) are shown in \textbf{A-C} and \textbf{D-F}, respectively for the W$_{53}$Ta$_{44}$V$_{3}$ and the W$_{53}$Ta$_{42}$V$_{5}$ RMEAs.}
		\label{fig:01}
	\end{figure}

Bright-Field TEM (BFTEM) micrographs of the W$_{53}$Ta$_{44}$V$_{3}$ and the W$_{53}$Ta$_{42}$V$_{5}$ recorded at three different conditions -- pristine, after annealing at 1173 K, and after irradiation at 1073 K -- are shown in Fig. \ref{fig:01}. Before both irradiation and annealing (Fig. \ref{fig:01}A and \ref{fig:01}D), these alloys exhibited equiaxed grains within the nanocrystalline regime (\textit{i.e.}, <100 nm). Apart from the chemical composition, no major differences between these two RMEAs were detected or identified. By analysing the BFTEM micrographs after annealing (Fig. \ref{fig:01}B and \ref{fig:01}E) and irradiation (Fig. \ref{fig:01}C and \ref{fig:01}F), it was noticeable that the W$_{53}$Ta$_{44}$V$_{3}$ RMEA experienced a modest grain growth, particularly noted after 20 dpa of irradiation. Conversely, the W$_{53}$Ta$_{42}$V$_{5}$ RMEA was seemingly unaltered after both annealing and irradiation. A quantitative analysis of average grain sizes as a function of the conditions studied in this work is shown in Table \ref{res:table01}.
	
These results exhibited in Fig. \ref{fig:01} are better comprehended considering the challenges on the application of nanocrystalline alloys in extreme environments. These challenges are intrinsically related with the (thermodynamic and morphological) stability of nanocrystalline materials \cite{cahn1990nanostructured,gleiter1992nanostructured,moriarty2001nanostructured} when subjected to exposure in both irradiation \cite{zhang2018radiation} and high-temperatures \cite{valiev2006principles}. In this context, the primary adverse consequence of such exposure is grain growth as numerous material properties stem from the average grain size. To date, and according to Zhang \textit{et al.} \cite{zhang2018radiation}, a limited number of studies addressed the grain size stability of nanocrystalline metallic alloys, and in this way, existing studies on this field are primarily focused on the behaviour of nanocrystalline elemental metals. Herein, it is consolidated that elemental metals in their nanocrystalline form experience significant grain growth under heating/annealing and accelerated grain growth under irradiation over a wide range of temperatures. With respect to nanocrystalline elemental metals exposed to extreme conditions, it is important to emphasize that theoretical models were already developed and experimentally validated to explain such observations \cite{alexander1991heat,alexander1991ion,alexander1993thermal,kaoumi2008thermal}. 

	\begin{table}[hb!]
		\centering
		\caption{Average grain size of RMEAs at pristine, after annealing and irradiation conditions.}
		\label{res:table01}
		\begin{tabular}{|c|cc|}
		\hline
		\multirow{2}{*}{\textbf{Alloy Condition}}                                        & \multicolumn{2}{c|}{\textbf{Average Grain Size {[}nm{]}}} \\ \cline{2-3} 
                                                                                 & \multicolumn{1}{c|}{53W-42Ta-3V [at.\%]}       & 53W-44Ta-5V [at.\%]      \\ \hline
		\textbf{\begin{tabular}[c]{@{}c@{}}Pristine\\ (0 dpa)\end{tabular}}              & \multicolumn{1}{c|}{19.2$\pm$0.9}                 & 25.9$\pm$1.0                \\ \hline
		\textbf{\begin{tabular}[c]{@{}c@{}}Annealed at 1173 K\\ (0 dpa)\end{tabular}}    & \multicolumn{1}{c|}{27.5$\pm$0.9}                 & 29.1$\pm$1.4                \\ \hline
		\textbf{\begin{tabular}[c]{@{}c@{}}Irradiated at 1073 K\\ (20 dpa)\end{tabular}} & \multicolumn{1}{c|}{29.4$\pm$1.3}                 & 27.3$\pm$2.1                \\ \hline
		\end{tabular}
	\end{table}
	
Nanocrystallinity allows achieving higher radiation tolerance within the scope of novel nuclear materials \cite{brailsford1976point,bullough1980sink,el2014ultrafine,aradi2020radiation}. This is due to the presence of a greater number of interfaces (namely grain boundaries) that enhances the capacity for radiation-induced crystalline defects to be absorbed and recombine effectively at these site-specific dependencies \cite{brailsford1976point,bullough1980sink}. This is historically known and defined as "sink efficiency", and for a metal, the sink efficiency increases exponentially with decreasing average grain size \cite{brailsford1976point,bullough1980sink}. When nanocrystalline alloys undergo grain growth or recrystallization due to energetic particle irradiation, their radiation tolerance is compromised. In such cases, the challenges that affect W for fusion reactors \cite{el2014situ,el2014ultrafine,el2017role,harrison2017engineering,el2018nanohardness,harrison2018damage,harrison2019use} also become relevant for both RHEAs and novel RMEAs. Concerning the current state of research on RHEAs, which are undergoing intensive development for fusion applications, it is worth noting that although there are limited studies on grain size stability under irradiation and high-temperature conditions, recent research reveals two critical findings: (i) not all nanocrystalline HEAs exhibit irradiation tolerance \cite{zhang2019thermal,tunes2021irradiation}, and (ii) certain RHEAs demonstrate superior performance under irradiation at high-temperatures than both coarse-grained and nanocrystalline W \cite{el2014situ,el2014ultrafine,el2017role,el2018nanohardness,el2019outstanding,el2021helium,el2023quinary}. 

Precipitation was observed in the quaternary nanocrystalline RHEA -- W$_{38}$Ta$_{36}$Cr$_{15}$V$_{11}$ \cite{el2019outstanding} -- when tested under heavy-ion irradiation, and although no significant alterations in the grain morphology were observed, such phase evolution effects are signs of degradation of the initially designed alloy. A new quinary RHEA \cite{el2023quinary} -- W$_{29}$Ta$_{42}$Cr$_{5}$V$_{16}$Hf$_{8}$ \cite{el2023quinary} -- was recently designed and synthesized. In terms of grain morphology after irradiation, the addition of Hf was associated with grain refinement/fragmentation during both annealing and irradiation, therefore indicating some grain instabilities that deserve further investigations. Although these previous works indicated that higher phase stability was achieved by increasing the chemical complexity, morphological changes were still observed in the microstructures of both alloys after irradiation. In comparison with both the W$_{38}$Ta$_{36}$Cr$_{15}$V$_{11}$ \cite{el2019outstanding} and the W$_{29}$Ta$_{42}$Cr$_{5}$V$_{16}$Hf$_{8}$ \cite{el2023quinary} RHEAs, the results presented in Fig. \ref{fig:01} show that new RMEAs with lower chemical complexity -- W$_{53}$Ta$_{42}$V$_{5}$ and the W$_{53}$Ta$_{44}$V$_{3}$ -- can be synthesized in its nanocrystalline form and still preserve high microstructural stability after both high-temperature annealing and irradiation exposure observed in the RHEAs with more alloying elements. The average grain sizes measured at pristine, after annealing at 1173 K and after irradiation at 1073 K and up to 20 dpa attests these findings. These results presented in Table \ref{res:table01} suggest that high-temperature annealing drives grain growth due to the removal of voided grain boundaries manifested as nanoporosity in the pristine samples after magnetron-sputtering deposition \cite{thornton1986microstructure}: evidence for such nanoporosity is presented in the supplemental information file. Conversely, irradiation at 1073 K after annealing indicates no detectable grain growth up to the tested dose of 20 dpa. Therefore, we can conclude that both W--Ta--V RMEAs exhibit a high degree of irradiation tolerance in terms of grain stability, as evidenced by the absence of growth and/or recrystallization after high-dose irradiation.
	
	\begin{figure}[hb!]
		\centering	
		\includegraphics[width=\textwidth]{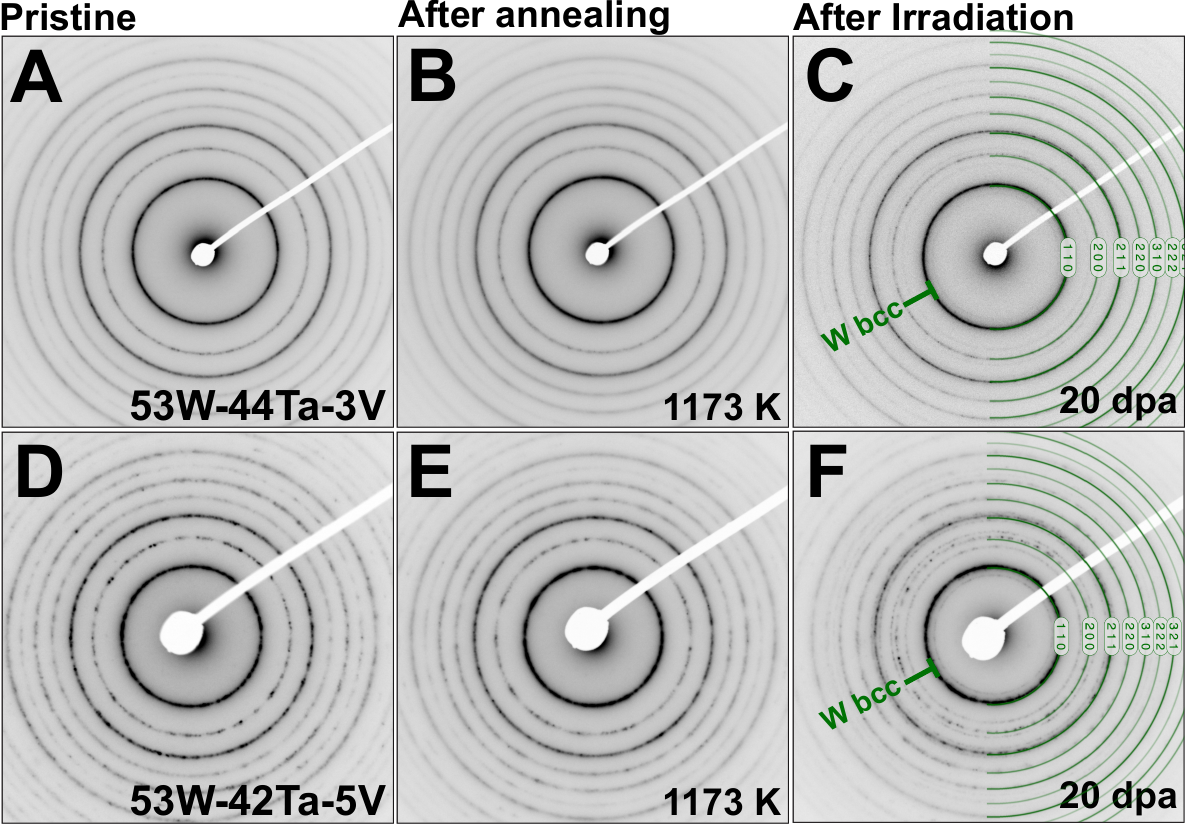}
		\caption{\textbf{Phase stability to annealing and irradiation} | SAED patterns collected at pristine condition, after annealing (at 1173 K) and after irradiation (at 1073 K) are shown in \textbf{A-C} and \textbf{D-F}, respectively for  W$_{53}$Ta$_{44}$V$_{3}$ and the the W$_{53}$Ta$_{42}$V$_{5}$ RMEAs. Additional rings were noted following irradiation, with a higher intensity observed in the W$_{53}$Ta$_{42}$V$_{5}$ RMEA (in \textbf{F}) compared to the W$_{53}$Ta$_{44}$V$_{3}$ RMEA (in \textbf{C}). Indexing was performed with data available in literature \cite{jette1935precision,noethling1925kristallstruktur,hellenbrandt2004inorganic,karen2005crystal}.}
		\label{fig:02}
	\end{figure}
	
The phase stability of both W--Ta--V RMEAs was also investigated with the SAED technique during the \textit{in situ} TEM experiments of high-temperature annealing and heavy-ion irradiation. Figs. \ref{fig:02}A and \ref{fig:02}D show the SAED patterns collected on the two different W--Ta--V RMEAs in their pristine condition; both alloys were indexed to have BCC crystalline structure matching W standards \cite{jette1935precision,hellenbrandt2004inorganic}. The phase of both alloys remained unaltered after high-temperature annealing, as noted in Figs. \ref{fig:02}B and \ref{fig:02}E. After irradiation up to 20 dpa, some additional Debye-Scherrer rings appeared in the SAED patterns of both alloys, as shown in Figs. \ref{fig:02}C and \ref{fig:02}F, although in the alloy with higher V content (Fig. \ref{fig:02}C), these rings are of lower intensity. These extra-rings could not be identified to any specific crystal structure or symmetry. It is important to emphasize that recently, these extra rings were also observed in the quinary W$_{29}$Ta$_{42}$Cr$_{5}$V$_{16}$Hf$_{8}$ RHEA after irradiation and as such, they were not indexed to any known phase \cite{el2023quinary}.
	
The microstructure of both W--Ta--V alloys, as depicted in Fig. \ref{fig:01}, reveals no phase transformation following high-temperature annealing or irradiation. The appearance of additional rings in the SAED patterns, as shown in Fig. \ref{fig:02}, after a dose of 20 dpa, suggests the potential occurrence of phase instabilities through precipitation, thus underscoring the need for further nanoscale post-irradiation screening with analytical microscopy methods.
	
\subsection{Chemical stability under extreme conditions}
\label{sec:discussion:nanochemistry}

\begin{figure}[hb!]
		\centering	
		\includegraphics[width=\textwidth]{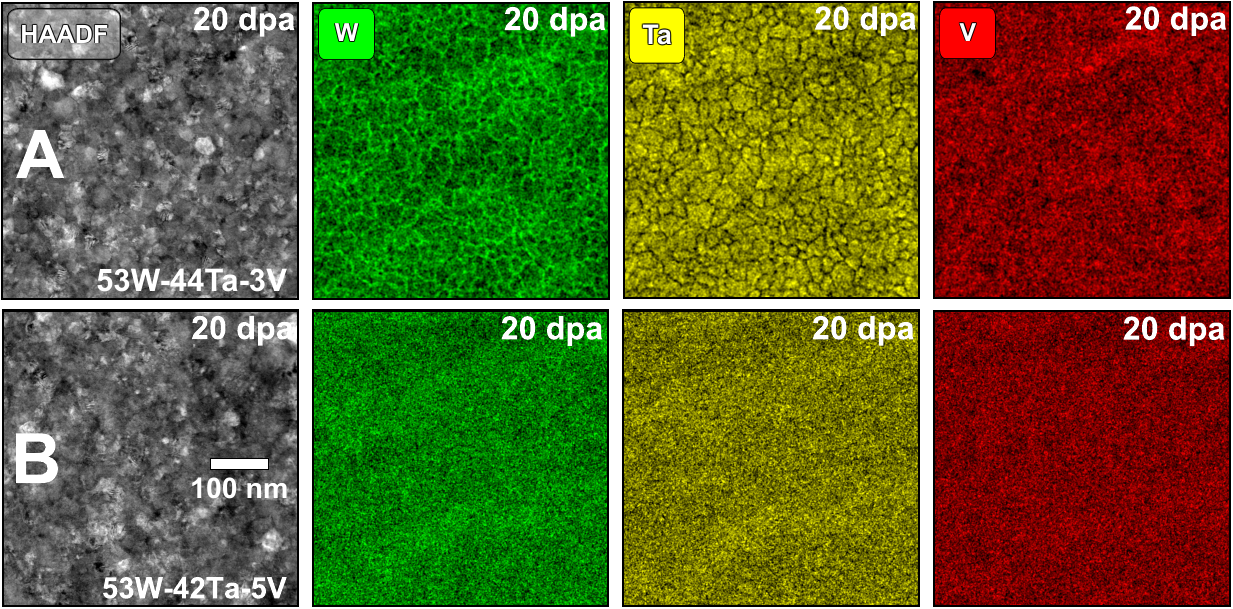}	
		\caption{\textbf{Nanoscale chemistry of the WTaV RMEAs after irradiation at 20 dpa} | High-magnification STEM-EDX mapping collected from both W$_{53}$Ta$_{42}$V$_{5}$ (top row) and the W$_{53}$Ta$_{44}$V$_{3}$ (bottom row) RMEAs reveal W segregation and Ta depletion at the nanocrystalline grain boundaries upon decreasing the V content in the W--Ta--V system. The W$_{53}$Ta$_{42}$V$_{5}$ RMEA neither exhibit radiation-induced segregation nor radiation-induced precipitation or phase instabilities/transformations, indicating a high-radiation tolerance at 20 dpa.}
		\label{fig:03}
	\end{figure}

\noindent Scanning Transmission Electron Microscopy coupled with Energy Dispersive X-ray (STEM-EDX) spectroscopy mapping was performed to assess the local chemistry of both W--Ta--V RMEAs after heavy-ion irradiation at 1173 K up to 20 dpa. Figs. \ref{fig:03}A and \ref{fig:03}B show the microstructure of the irradiated W$_{53}$Ta$_{42}$V$_{5}$ and the W$_{53}$Ta$_{44}$V$_{3}$ RMEAs, respectively, using both high-angle annular dark-field (HAADF) and EDX elemental maps. While the W$_{53}$Ta$_{44}$V$_{3}$ RMEA clearly indicates the occurrence of W segregation and Ta depletion at grain boundaries, no such segregation was observed or detected in the W$_{53}$Ta$_{42}$V$_{5}$ RMEA. Complimentary STEM-EDX maps for both pristine and annealed-only conditions are shown in the supplemental information file.
	
The nanoscale analytical assessment presented in Fig. \ref{fig:03} for both alloys after irradiation suggests that lower concentrations of V reduce the chemical stability of the W--Ta--V system. Interestingly, the W$_{53}$Ta$_{42}$V$_{5}$ RMEA not only exhibit high morphological stability under irradiation (Fig. \ref{fig:02}), preventing grain growth or recrystallization up to a dose of 20 dpa at 1073 K, but also presents high chemical stability given the absence of grain boundary segregation under the irradiation conditions studied. These results shed light on the behavior that these RMEAs -- manifested by the reduction on the number of alloying elements when compared with conventional RHEAs \cite{el2019outstanding,el2023quinary} -- can have their radiation response tailored as a function of the concentration of V. Therefore, further investigation is needed to explore the specific role of V in the context of the W--Ta--V system.
	
\subsection{The role of V in the stability and radiation response of chemically simplified RMEAs}
\label{sec:discussion:roleofV}

\begin{figure}[hb!]
	\centering	
	\includegraphics[width=\textwidth]{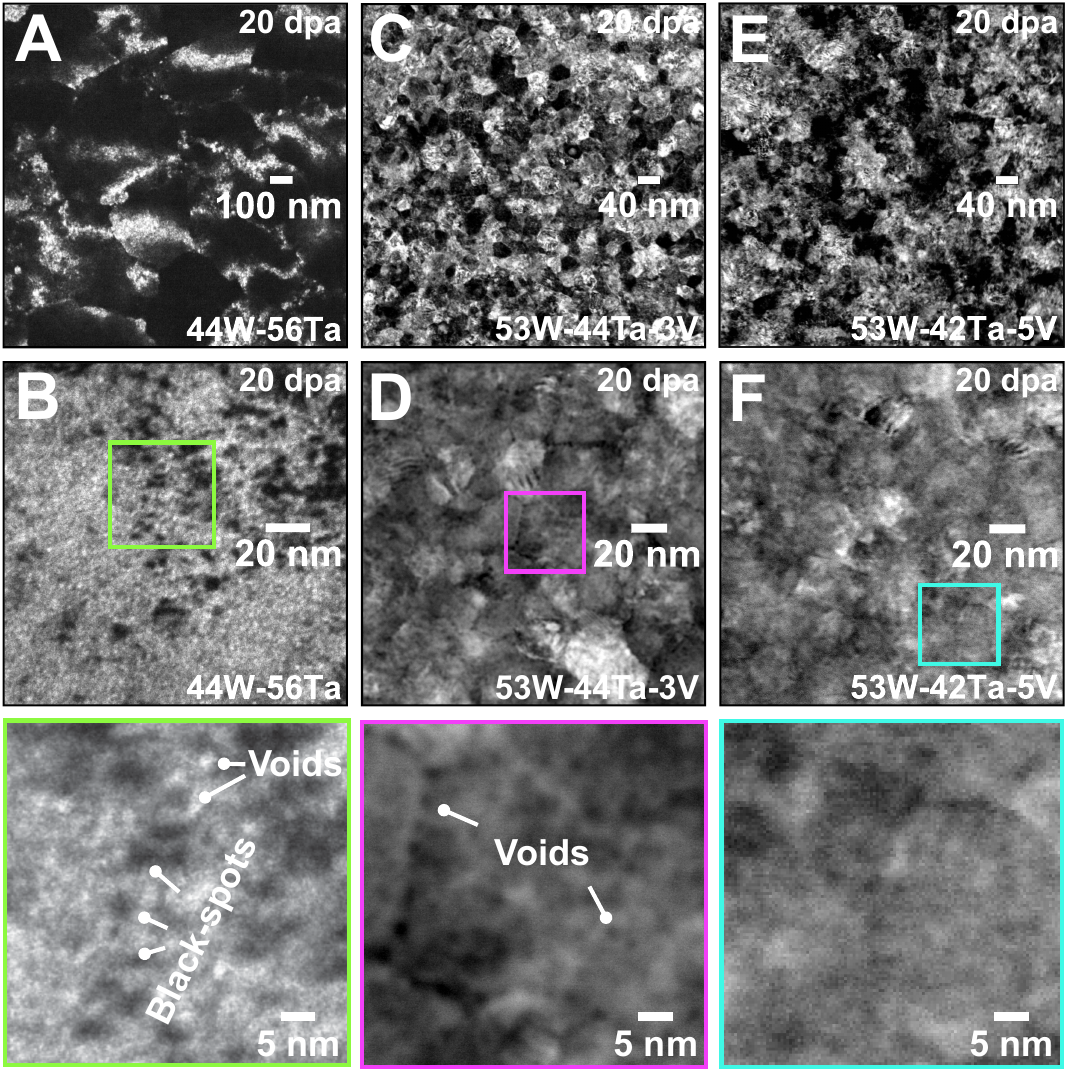}
	\caption{\textbf{Comparison between WTa and WTaV RMEAs at 20 dpa} | Grain stability and defects after irradiation are observed in the UFG W$_{44}$Ta$_{56}$Ta alloy as shown in the DFTEM micrograph in \textbf{A} and the underfocused BFTEM micrograph in \textbf{B}: this binary alloy exhibit both voids and black-spots as particularly shown in \textbf{B} and its green-square inset. Grain morphology for both irradiated W$_{53}$Ta$_{44}$V$_{3}$ and W$_{53}$Ta$_{42}$V$_{5}$ RMEAs are shown in the DFTEM micrographs in \textbf{C} and \textbf{E}: both alloys' microstructures are stable in terms of grain-size after 20 dpa. HAADF was used to better resolve voids in the RMEAs as shown in micrographs \textbf{D} and \textbf{F}, but voids were clearly resolvable in the W$_{53}$Ta$_{44}$V$_{3}$ RMEA, particularly noted in its pink-square inset, whereas in the W$_{53}$Ta$_{42}$V$_{5}$ RMEAs, voids were barely resolvable, particularly noted in its cyan-square inset. Note: Image in \textbf{B} is a underfocused BFTEM micrograph where voids appear ``round and white'', whereas micrographs \textbf{D and F} are HAADF where voids appear ``round-black''.}
	\label{fig:04}
\end{figure}

\noindent To investigate the role of V on the radiation response of the W--Ta--V system, additional high-temperature irradiation experiments were conducted using a binary UFG W$_{44}$Ta$_{56}$ alloy synthesized under similar deposition conditions. The microstructure of the UFG W$_{44}$Ta$_{56}$ alloy, as observed through both dark-field (DFTEM) and BFTEM after 20 dpa, is depicted in Figs. \ref{fig:04}A and \ref{fig:04}B, respectively.

Following a heavy-ion irradiation of 20 dpa, this binary alloy exhibited extensive nucleation of voids and black spots, as particularly evident in the BFTEM micrograph presented in Fig. \ref{fig:04}B. It is worth emphasizing that Yi \textit{et al.} reported an extensive chain of irradiation-induced defects to nucleate and evolve in a coarse-grained binary alloy W-5Ta (in wt.\%), a terminal solid solution, at doses as low as 1.2 dpa \cite{yi2019high}. These past results from Yi \textit{et al.} \cite{yi2019high} emphasize that regardless of grain-size (coarse- or ultrafine-grained) or composition (highly concentrated or terminal solid solution), WTa binary alloys suffer with radiation damage at low doses. Similarly to the irradiated UFG W$_{44}$Ta$_{56}$Ta, the microstructures of both the W$_{53}$Ta$_{42}$V$_{5}$ and the W$_{53}$Ta$_{44}$V$_{3}$ RMEAs after the same 20 dpa heavy-ion irradiation are shown in Fig. \ref{fig:04}(C,D) and \ref{fig:04}(E,F), respectively, utilizing DFTEM and HAADF imaging. It is important to emphasize that the binary alloy as-synthesized has larger grain sizes (pertaining to the UFG regime) when compared with the ternary RMEAs, which are nanocrystalline. Instead of BFTEM imaging, HAADF was preferred to investigate the presence of voids in both ternary RMEAs as sparse voids are directly viewed as ``round-black''features using this imaging mode.

While some voids were observed within the intra-granular regions of the W$_{53}$Ta$_{44}$V$_{3}$ RMEA (Fig. \ref{fig:04}E), voids were not clearly detected or were not resolvable in the W$_{53}$Ta$_{42}$V$_{5}$ RMEA when imaged under identical conditions. It is important emphasizing that the observation of voids in both alloys is dependent on the spatial resolution of our electron-microscopes which is around 0.25 nm. In addition, the results indicate that nanocrystalline RMEAs exhibit superior performance compared to UFG binary alloy in terms of their resistance to the formation of black-spots and voids under irradiation, directly suggesting that the small additions of element V, increases significantly the radiation response of the binary alloy. However, it is important noting that in the specific conditions of our study, nanocrystallinity was only achieved in the W--Ta system when V was added as an alloying element. This observation underscores the potential role of V in influencing the radiation response of these new RMEAs, warranting further investigation.

\begin{table}[hb!]
\begin{center}
\caption{Average void size (diameter) and areal densities for all samples after irradiation.}
\label{res:table02}
\begin{tabular}{|l|l|l|}
\hline
\textbf{Irradiated Alloy (20 dpa at 1073 K)} & \textbf{Average Void Size {[}nm{]}} & \textbf{Average Areal Density {[}\#$\cdot$nm$^{-2}${]}} \\ \hline
W$_{44}$Ta$_{56}$                            & 1.9$\pm$0.1                       & 8.9$\pm$0.2$\times$10$^{-3}$                          \\ \hline
W$_{53}$Ta$_{44}$V$_{3}$                     & 1.7$\pm$0.1                       & 7.0$\pm$1.1$\times$10$^{-3}$                           \\ \hline
W$_{53}$Ta$_{42}$V$_{5}$                     & Not observed/detected               & Not observed/detected                          \\ \hline
\end{tabular}
\end{center}
\end{table}

A quantification of both voids and areal density of voids is presented in Table \ref{res:table02}. Based on our observations of the studied alloys irradiated to 20 dpa, the W$_{53}$Ta$_{44}$V$_{3}$ RMEA exhibits the smaller average void sizes and areal density than the binary UFG W$_{44}$Ta$_{56}$ alloy: already an indicative of higher radiation tolerance for the ternary alloy with V when compared with the binary alloy. However, the most interesting observation arising from this work is that upon increasing the V content from 3 to 5 at.\%, voids were not observed in the W$_{53}$Ta$_{42}$V$_{5}$ RMEA. The radiation tolerance of these new RMEAs is better evaluated when recent data on \textit{in situ} TEM ion irradiation of relevant fusion materials is taken into consideration. Under similar irradiation temperatures and doses to this present work, \textit{in situ} TEM with He implantation revealed He bubbles with an average diameter of 6.6 nm and an areal density of 2.7$\times$10$^{-2}$ bubbles$\cdot$nm$^{-2}$ for bulk W \cite{el2020revealing,el2020temperature}. For UFG W, the average diameter of bubbles was reported to be 6.4 nm with an areal density of 7.5$\times$10$^{-2}$ bubbles$\cdot$nm$^{-2}$ \cite{el2020situ}. The previous W$_{38}$Ta$_{36}$Cr$_{15}$V$_{11}$ RHEA exhibited He bubbles of around 2.2 nm in diameter with an areal density of 1.5$\times$10$^{-1}$ bubbles$\cdot$nm$^{-2}$ \cite{el2019outstanding}. Both average void size and areal density for the W$_{53}$Ta$_{44}$V$_{3}$ RMEA investigated in this work are smaller than He bubbles in bulk W \cite{el2020revealing,el2020temperature}, UFG W \cite{el2020situ} and previous chemically-complex quaternary RHEAs \cite{el2019outstanding}, together with the fact that the W$_{53}$Ta$_{42}$V$_{5}$ RMEA exhibited no detectable voids. The results in this work are emphasising the new potential of such RMEAs within the ternary system of W--Ta--V present to be considered candidate fusion materials. 

\subsection{Chemical short-range order within the W--Ta--V system}
\label{sec:discussion:sroinwtav}

\noindent To gain a deeper understanding of the important role of V on radiation-induced stability in the considered RMEAs, an atomistic Monte-Carlo (AMC) modelling approach based on density functional theory (DFT) and cluster expansion Hamiltonian (CEH) methods, developed recently for multi-component system \cite{ducnm2021,AFC2019,damian2020,andrew2023}, has been employed to predict the chemical short-range order (CSRO) of the three different pairs (W-Ta, W-V and Ta-V) and thermodynamic properties of both W$_{53}$Ta$_{44}$V$_{3}$ and  W$_{53}$Ta$_{42}$V$_{5}$ alloys.

	\begin{figure}[hb!]
		\centering	
		\includegraphics[width=0.72\textwidth]{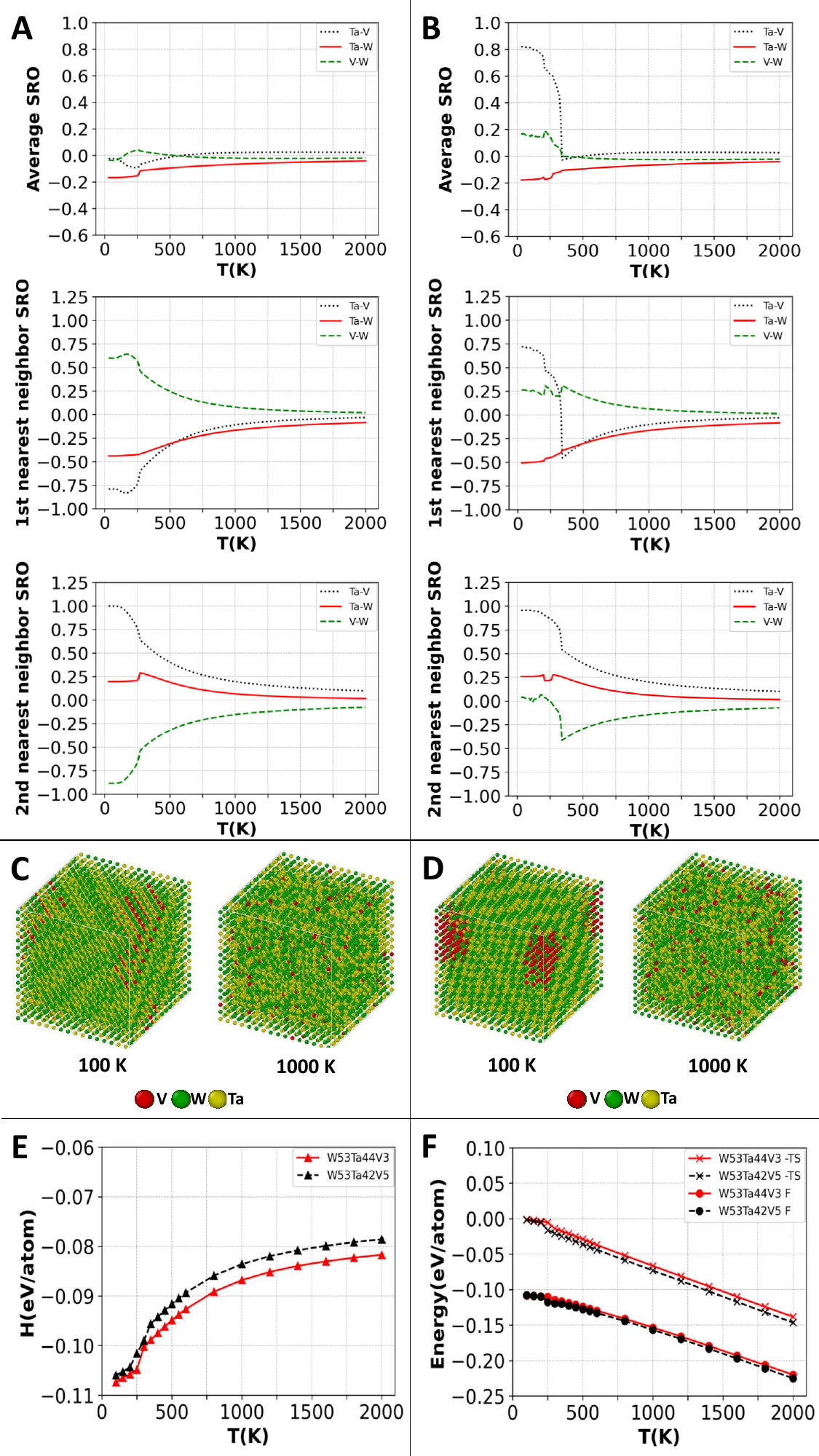}
		\caption{\textbf{DFT-CEH based AMC simulations for W$_{53}$Ta$_{44}$V$_{3}$ and W$_{53}$Ta$_{42}$V$_{5}$} | The predicted Warren-Cowley chemical short-range order parameters (average, 1NN and 2NN) for different pairs of atoms (W-Ta: blue; W-V: red and Ta-V; yellow) as a function of temperature for W$_{53}$Ta$_{44}$V$_{3}$ (A) and W$_{53}$Ta$_{42}$V$_{5}$ (B). Atomic structures obtained from AMC simulations at 100K and 1000K are shown for W$_{53}$Ta$_{42}$V$_{3}$ (C) and W$_{53}$Ta$_{44}$V$_{5}$ (D). The enthalpy of mixing (E), configuration entropy and free energy (F) calculated as a function of temperature for the two considered RMEA alloys. }
		\label{fig:05}
	\end{figure}

In agreement with our previous investigations in quinary alloys containing W, Ta and V \cite{damian2020,andrew2023}, the calculated Warren-Cowley CSRO parameter of the W-Ta pair averaged over first nearest-neighbour (1NN) and second nearest-neighbour (2NN) in the BCC system is negative as a function of temperature in both W$_{53}$Ta$_{44}$V$_{3}$ and W$_{53}$Ta$_{42}$V$_{5}$ as shown in Figs. \ref{fig:05}A and \ref{fig:05}B, respectively. This common behaviour resulting from the dominance of chemical bonding between W and Ta atoms in the 1NN and the repulsion at the 2NN can be explained by the negative enthalpy of mixing, which has been predicted in all ranges of composition in the binary W--Ta system \cite{muzyk2011}. In particular, at the composition closer to equiatomic, the above finding is related to the existence of a B$_{2}$ phase in a simple interpretation of the binary W--Ta phase diagram \cite{turchi2001} whereas more accurate DFT calculations predicted the stability of an orthorhombic A$_{6}$B$_{6}$ phase \cite{muzyk2011}.

The average CSRO of W-V pair in the two ternary alloys with rich W and Ta concentration is, however, quite different to those of W-Ta as depicted in Figs. \ref{fig:05}A and \ref{fig:05}B although a similar trend of negative enthalpy of mixing has also predicted in the binary W-V system \cite{muzyk2011}. Due to low V concentration, the CSRO W-V pair is found to be strongly positive for the 1NN, while it is mostly negative for the 2NN as a function of temperature. There is, however, a significant difference between the two alloys with different V concentrations. Namely, at low temperatures (T < 250K), the 2NN CSRO between W-V changes from negative to positive for the W$_{53}$Ta$_{42}$V$_{5}$ RMEA, whereas it continues to be negative for the W$_{53}$Ta$_{44}$V$_{3}$ RMEA. This means that for the latter case, V atoms are strongly to be present in the 2NN shell of a W atom, while in the former case they tend to remain far from W. 

In a consistency with the above analysis for the 2NN CSRO of W-V, the second interesting difference between the two alloys considered W$_{53}$Ta$_{42}$V$_{5}$ and W$_{53}$Ta$_{44}$V$_{3}$ is found in the 1NN CSRO of Ta-V pairs. As both Ta and V are transition metals in group 5 of the periodic table, the enthalpy of mixing is expected to be positive in the binary Ta-V system and accordingly the 1NN CSRO should be positive indicating the segregation tendency between Ta and V. In a strong variance with the binary alloy system, the CSRO of the 1NN Ta-V pair in the W$_{53}$Ta$_{44}$V$_{3}$ RMEA is negative for all temperatures demonstrating the important effect of W concentration on the chemical SRO between Ta and V. In the W$_{53}$Ta$_{42}$V$_{5}$ RMEA, a similar negative trend for the 1NN CSRO of Ta-V pair is observed for temperatures higher than about 350 K, but it becomes positive for T < 350K. This dramatic change results in different behavior of the average CSRO parameter for Ta-V of  W$_{53}$Ta$_{42}$V$_{5}$ in comparison with those of W$_{53}$Ta$_{44}$V$_{3}$ at low temperature. It is worth mentioning that our comparison of the predicted CSRO at low-temperature for the 1NN Ta-V pairs and those of the 2NN for W-V pairs in the two considered RMEAs serves as a crucial precursor to differentiate their properties systematically as a function of temperatures including the at high temperatures. For instance, at T = 1200 K, our calculated CSRO for 1NN Ta-V and 2NN W-V are found to be -0.0803$\pm$0.0005  and -0.1283$\pm$0.0005, respectively, in W$_{53}$Ta$_{44}$V$_{3}$ that are stronger negative than the corresponding values of -0.0736$\pm$0.0006 and -0.1194$\pm$0.0003, respectively, found in the RMEA with higher V, \textit{i.e.}, the W$_{53}$Ta$_{42}$V$_{5}$ . Stronger negative CSRO at higher temperatures (and irradiation as it enhances solid-state diffusion) suggest a tendency for the RMEA to exhibit segregation as observed in the W$_{53}$Ta$_{44}$V$_{3}$ RMEA (see Fig. \ref{fig:03}B). This segregation is suppressed upon minor additions of V to compose the W$_{53}$Ta$_{42}$V$_{5}$ RMEA, which has a decreased CSRO effect, resulting in increasing the chemical stability of the random solid solution.

The above analysis of CSRO parameters demonstrates a sensitivity of microstructural stability with respect to the CSRO as a function of both V concentration and temperature in the W--Ta--V system. Figures \ref{fig:05}C and \ref{fig:05}D show the results of AMC simulations in a cell with $16.000$ atoms obtained at 100K and 1000K for W$_{53}$Ta$_{44}$V$_{3}$ and W$_{53}$Ta$_{42}$V$_{5}$ RMEAs, respectively. A stronger tendency of ordering surrounding V atoms for the former system was in good agreement  with the SAED indexing in W$_{53}$Ta$_{44}$V$_{3}$ after irradiation (see Fig. \ref{fig:02}F). It is worth emphasising that at high temperatures, the 1NN contribution to the CSRO behaviour for V-Ta pair is dominantly negative and much stronger for W$_{53}$Ta$_{44}$V$_{3}$ than those of W$_{53}$Ta$_{42}$V$_{5}$ as it can be seen from \ref{fig:05}A and \ref{fig:05}B. Our AMC simulations also indicate the more homogeneous atomic distribution in W$_{53}$Ta$_{42}$V$_{5}$, where There are no observations of radiation induced defects (voids) at 20 dpa at high temperature as shown in Table. \ref{res:table02}.Finally, our calculations of enthalpy of mixing (Figs \ref{fig:05}E), configurational entropy and free energy depicted in Figs \ref{fig:05}F showed a more negative enthalpy of mixing for W$_{53}$Ta$_{44}$V$_{3}$ over W$_{53}$Ta$_{42}$V$_{5}$ over all the temperature range mainly due to V-driven chemical ordering effect in the former RMEA.

It is important to establish the correlation between V-Ta CSRO with the observed microstructural changes after irradiation for the two W--Ta--V RMEAs with respect to a small variation of V concentration from our AMC simulations. The lower-V alloy W$_{53}$Ta$_{44}$V$_{3}$ has the radiation-induced second V-bcc phase due to anomalous negative average CSRO parameter between V-Ta while the higher-V alloy W$_{53}$Ta$_{42}$V$_{5}$  has a dominant single W-bcc phase with a segregation tendency between V and Ta associated with the positive average CSRO. It is worth emphasizing the important role of V on distinct and low radiation-induced defect formation properties in comparison with those of W and Ta \cite{dnm2006,dnd2007} and therefore providing insight into enhanced radiation resistance in the bcc MEAs W--Ta--V system. Crucially, the prediction of strong change for Ta-V CSRO properties between W$_{53}$Ta$_{44}$V$_{3}$ and W$_{53}$Ta$_{42}$V$_{5}$ provides an excellent support to the experimental observation of a systematic suppression of voids from the binary W$_{44}$Ta$_{56}$  to the RMEAs as it is shown in Table \ref{res:table02}. According to molecular dynamic (MD) simulations, a slow self-interstitial atom (SIA) and fast vacancy diffusion due to weak binding energies of V dumbbells in comparison with significantly larger ones of W and Ta lead to the closer mobilities of vacancies and SIAs, which enhances defect recombination in the annealing process in W--Ta--V alloys \cite{Li2023}. To support these findings, additional AMC simulations have been performed for W$_{44}$Ta$_{56}$ with 0.5$\%$ vacancy using the DFT-CEH, as reported elsewhere \cite{ducnm2021}. A strong presence of voids for this binary system is observed and in good agreement with the experimental observation shown in Fig. \ref{fig:04}D for W$_{44}$Ta$_{56}$ (see Fig. S1 from Supplementary Information). Importantly, the average void size obtained from AMC simulations equal to 1.5 $\pm$0.4 nm agrees with the value from the experiment reported in Table \ref{res:table02}. A further systematic investigation will be focusing on the defect properties as a function of V alloying concentration in radiation-induced microstructure stability of W--Ta--V system. Finally, it is interesting to predict new candidate MEAs with enhancing radiation resistance from a CSRO behaviour similar to those of  W--Ta--V RMEAs investigated in this work. One of the candidates would be W--Ti--V alloys for which our AMC simulations were performed based on CEH developed in \cite{damian2020}. It is found that for W$_{53}$Ti$_{42}$V$_{5}$ the temperature dependence of CSRO between Ti-V as well as the enthalpy of mixing are similar to those of Ta-V at the temperature around 250K for W$_{53}$Ta$_{42}$V$_{5}$ leading to a potentially single bcc-based W phase at high-temperature region  (See Fig. S2 from Supplementary Information).
 
\section{Conclusions}
\label{sec:conclusions}

\noindent We have demonstrated by means of irradiation-testing and characterization experiments that new RMEAs can be developed and their radiation response can be tailored as a function of minor modifications in the alloys' chemistry. Two new nanocrystalline RMEAs were synthesized and investigated in this work: W$_{53}$Ta$_{42}$V$_{5}$ and W$_{53}$Ta$_{44}$V$_{3}$. Atomistic simulations of short-range order in these bcc RMEAs unravelled the significant role of V alloying effects on radiation-induced micro-structural stability and defect formation properties in W--Ta--V via a systematic investigation of dependence in CSRO for the Ta-V in 1NN and those for the W-V in 2NN as a function of V concentration and temperature. Minor variations in V content lead to differences in phase stability and segregation behavior that were experimentally observed and computationally validated. The transition from RHEAs to RMEAs has the potential to greatly influence the quest for materials with high radiation tolerance for the practical application in future nuclear fusion reactors. In addition, the WTaV RMEAs studied in this work address the concerns of materials' activation whilst under in-service in a fusion reactor: they both present reduced activation when comparing with existing RHEAs so far investigated. This potential arises from the reduction in chemical complexity, achieved by decreasing the number of alloying elements without jeopardizing the high-stability inherent to HEAs with four or more constituents. This simplification is aimed at facilitating the metallurgical processes for synthesizing these alloys in bulk and at larger scales, which is a necessary step for achieving both technological readiness and commercialisation of materials for future fusion reactors. Therefore, this study lays the groundwork for future investigations into defect properties as a function of V alloying concentration and the exploration of new candidate multi-element alloys with enhanced radiation resistance and simplified-chemistry, the latter manifested by a reduction on number of alloying elements, thus deviating from the original high-entropy alloy concept \cite{cantor2002novel} to a new and optimized way to think and design the future materials for nuclear fusion.
	
\section{Materials and Methods}
\label{sec:matmet}
	
\subsection{Synthesis of the WTa and WTaV alloys}
\label{sec:matmet:synthesis}
	
\noindent Metallic alloys in the binary system W--Ta and the ternary system W--Ta--V were prepared in a form of thin solid films using the magnetron-sputtering deposition technique from elemental targets with nominal 99.99 at.\% purity. The deposition was performed at room temperature, 3 mTorr of pressure and with no bias voltage. Multiple thin solid films were deposited on pure NaCl (100) substrates from Hilger Crystals, Inc. After deposition, the substrates were dissolved in a deionized water and pure ethanol solution and the floating films were placed onto pure Mo mesh grids for TEM analysis and irradiation. The thickness of the deposited alloys was estimated to be $\approx$ 70 nm using profilometry. .
	
\subsection{Heavy-ion irradiations with \textit{in situ} TEM}
\label{sec:matmet:irradiation}
	
\noindent All the alloys investigated in this work were irradiated with 1 MeV Kr heavy-ions with \textit{in situ} TEM at the Intermediate Voltage Electron Microscope (IVEM) Tandem facility located at the Argonne National Laboratory. Prior irradiation, the samples were subjected to 10 min of annealing at 1173 K using a Gatan double-tilt heating holder (furnace-type). Irradiations were then carried out at temperature of 1073 K with a Kr flux measured to be 1.25$\times$10$^{12}$ ions$\cdot$cm$^{-2}\cdot$s$^{-1}$. BFTEM micrographs and SAED patterns were recorded before and after irradiation. During irradiation, the samples were under constant electron-beam monitoring using a Hitachi TEM 9000 operating a LaB$_6$ filament at 300 keV. Multiple videos were recorded during both annealing and irradiation and can be found in the supplemental files. Fluence-to-dpa calculations were performed using the Stopping and Range of Ions in Matter (SRIM) Monte Carlo code version 2013 \cite{ziegler2010srim}. For these calculations, a procedure described by Stoller \textit{et al.} was used \cite{stoller2013use}, which consists in using the Quick-Damage calculation mode setting the displacement energies for all transition metals to 40 eV. The alloys in this study were subjected to 1 h of irradiation, achieving a fluence of 4.50$\times$10$^{15}$ ions$\cdot$cm$^{-2}$, equivalent to an average of 20 dpa across the assumed 100 nm sample thicknesses and the SRIM calculated default densities (in g$\cdot$cm$^{-3}$). A plot of dose in dpa as a function of the irradiation depth is shown in the supplemental information file.
	
\subsection{Post-irradiation screening with conventional and analytical S/TEM}
\label{sec:matmet:postirradiation}
	
\noindent Following irradiation, the samples were allowed to cool within the TEM's vacuum environment by deactivating the external heating supply provided by GATAN. After cooling, these samples were removed for further post-irradiation carried out at a FEI Titan 80-300 STEM located at the Los Alamos National Laboratory and also using a Thermo Fisher Talos F200X located at the Montanuniversit\"at Leoben. Post-irradiation characterisation was performed using STEM mode and the High-Angle Annular Dark-Field (HAADF) and Bright-Field (BF) detectors as well as Energy Dispersive X-ray (EDX) spectroscopy that was used for elemental mapping of the unirradiated and irradiated alloys. Dark-Field TEM (DFTEM) was also used. Selected-Area Electron Diffraction (SAED) was used to assess the phase stability of the alloys in pristine condition as well as after annealing at 1173 K and after irradiation at 1073 K. SAED pattern indexing was performed with data available in the ICSD database \cite{hellenbrandt2004inorganic}, more specifically: ICSD-52268 for W (BCC) \cite{jette1935precision}, ICSD-53786 for Hf (HCP) \cite{noethling1925kristallstruktur}, and ICSD-171003 for V (BCC) \cite{karen2005crystal}.
	
\subsection{Modelling and calculation techniques}
\label{sec:matmet:modeling}
	
\noindent In this study, Atomistic Monte Carlo (AMC) simulations were performed using Cluster Expansion Hamiltonian (CEH) formalism which has been recently developed for studying the phase stability of compositionally complex and multi-component systems as a function of temperature and irradiation damage for different structural materials based on W-based \cite{ducnm2021,muzyk2011,jan2017,damian2020,andrew2023} and Fe-based \cite{jan2015,mark2020}alloys. The CEH can be written in the following formula \cite{ducnm2021}:
 
\begin{equation}
{\Delta H}_{mix}^{CEH}( \vec{\sigma } )=
\sum\limits _{\omega,n,s} J^{(s)}_{\omega,n} m^{(s)}_{\omega,n}  \langle {\Gamma_{\omega',n'}^{(s')}(\vec{\sigma})}\rangle_{\omega,n,s}, 
\label{eq:CEH}
\end{equation}

where the summation is performed over all the clusters, distinct under symmetry operations in the BCC lattice for the present study. $\vec{\sigma}$ denotes the ensemble of occupation variables in the lattice. $\omega$ and $n$ are the cluster size and its shell label, respectively.
$m^{(s)}_{\omega,n}$ denotes the site multiplicity of the decorated clusters (in per-lattice-site units); and $J^{(s)}_{\omega,n}$ represents the many-body effective cluster interaction (ECI) energy corresponding to the same $(s)$ decorated cluster. In Eq.(\ref{eq:CEH}),  $\langle {\Gamma_{\omega',n'}^{(s')}(\vec{\sigma})}\rangle_{\omega,n,s}$ denotes the cluster function, averaged over all the clusters of size, $\omega'$, and label, $n'$, decorated by the sequence of point functions,$(s')$. Within the matrix formulation of CEH for a system with $K$ elements, the cluster function is related to the probability function of finding cluster via the formula \cite{AFC2019}: 

\begin{equation}
y_{\omega,n}^{(AB\cdots)}={\overbrace{(\bar{\bar{\tau}}^{-1}_{K}\otimes\cdots\otimes\bar{\bar{\tau}}^{-1}_{K})}^{\omega}}_{AB\cdots,ij\cdots} {\langle \Gamma_{\omega,n}^{(ij\cdots)}\rangle}
\label{eq:clusterprob}
\end{equation}

where the matrix elements of the inverse of the $(\bar{\bar{\tau}}_{K})$ matrix are defined by the expression \cite{AFC2019}:

\begin{equation}
(\bar{\bar{\tau}}^{-1}_{K})_{ij}=\begin{cases}
\frac{1}{K} & \textrm{ if }j=0\textrm{ }, \\
-\frac{2}{K} \cos\left(2\pi\lceil\frac{j}{2}\rceil\frac{\sigma_i}{K}\right) & \textrm{ if }j>0\textrm{ and }j-1<K\textrm{ and j odd},  \\
-\frac{2}{K} \sin\left(2\pi\lceil\frac
{j}{2}\rceil\frac{\sigma_i}{K}\right) & \textrm{ if }j>0\textrm{ and j even}, \\
-\frac{1}{K} \cos\left(2\pi\lceil\frac{j}{2}\rceil\frac{\sigma_i}{K}\right) & \textrm{ if }j-1=K\textrm{ and j odd}.
\end{cases}
\label{eq:tauinverse}
\end{equation}

and $i,j = 0,1,2,...(K-1)$, $j$ and $\lceil\frac{j}{2}\rceil$ stands for the ceiling function - rounding up to the closest integer. From the Eq.(\ref{eq:clusterprob}), the point probability function is written as:
\begin{equation}
y^{A}_{1,1}=\sum\limits _{s} (\bar{\bar{\tau}}^{-1}_{K})_{A,(s)} \langle \Gamma_{1,1}^{(s)}\rangle,
\label{eq:point-probability}
\end{equation}
and the pair probability function is determined by the following formula :
\begin{equation}
y_{2,n}^{AB}= \sum\limits _{s} (\bar{\bar{\tau}}^{-1}_{K}\otimes\bar{\bar{\tau}}^{-1}_{K})_{A,B,(s)} \langle \Gamma_{2,n}^{(s)}\rangle.
\label{eq:pair-probability}
\end{equation}

Eq.(\ref{eq:pair-probability}) allows the two-body cluster probability to be linked with the Warren-Cowley short-range order (SRO) parameter, $\alpha_{2,n}^{(AB)}$,  via the definition \cite{warren1990,cowley1965}:  

\begin{equation}
y_{2,n}^{AB}=x_{A} x_{B} ( \, 1-\alpha_{2,n}^{AB} ) \, 
\label{eq:WC}
\end{equation}

where $x_{A}$ and $x_{B}$ denote the bulk concentration of the chemical species A and B, respectively. In the case where  $\alpha_{\text{2,n}}^{\text{AB}}=0$, the pair probability is given by the product of their concentrations $\text{x}_{\text{B}}\text{x}_{\text{B}}$ corresponding to random configuration of A and B species in an alloy system.  In the case of $\alpha_{\text{2,n}}^{\text{AB}}>0$, clustering or segregation between A-A and B-B pairs is favoured and for	$\alpha_{\text{2,n}}^{\text{AB}}<0$, the chemical ordering of A-B pairs occurs. By combining Eqs.(\ref{eq:point-probability}),(\ref{eq:pair-probability} and (\ref{eq:WC}), the SRO parameters for $K$-component system can be calculated by the general expression as follows \cite{ducnm2021}:

\begin{equation}
\alpha_{2,n}^{AB} = 1- \dfrac{\sum\limits _{s} (\bar{\bar{\tau}}^{-1}_{K}\otimes\bar{\bar{\tau}}^{-1}_{K})_{A,B,(s)} \langle \Gamma_{2,n}^{(s)}\rangle}{\big( \sum\limits _{s} (\bar{\bar{\tau}}^{-1}_{K})_{A,(s)} \langle \Gamma_{1,1}^{(s)}\rangle \big) \big( \sum\limits _{s} (\bar{\bar{\tau}}^{-1}_{K})_{B,(s)} \langle \Gamma_{1,1}^{(s)}\rangle \big)}.
\label{eq:sromatrix}
\end{equation}

The expression to calculate the average SRO parameter for first and second nearest neighbours in a BCC lattice  is given by \cite{Mirebeau2010}:

\begin{equation}
\alpha_{avg}^{AB}=\frac{8\alpha_{2,1}^{AB}+6\alpha_{2,2}^{AB}}{14}
\label{eq:SRO_avg}
\end{equation}

\noindent where $\alpha_{2,1}^{AB}$ and $\alpha_{2,2}^{AB}$  are the first and second nearest neighbours SRO parameters, respectively.

The CEH, defined by Eq. (\ref{eq:CEH}) can be used to explicitly determine the configuration entropy of a $K$-component system via the thermodynamic integration method \cite{jan2015,mark2020}.  Here the entropy is computed from fluctuations of the enthalpy of mixing at a given temperature using the following formula

\begin{equation}
S_{conf}[T]=\int_{0}^{T} \dfrac{C_{conf}(T')}{T'} dT'=\int_{0}^{T} \dfrac{\langle [{\Delta H_{mix}^{CEH}(T')}]^{2}\rangle-{\langle [\Delta H_{mix}^{CEH}(T')]\rangle}^{2}}{{T'}^{3}} dT',
\label{eq:TDI}
\end{equation}
where ${\langle [\Delta H_{mix}^{CEH}(T')]^{2}\rangle}$ and $\langle [\Delta H_{mix}^{CE}(T')]\rangle^{2}$ are the square of the mean and mean square enthalpies of mixing, respectively. The statistical average over configurations at finite temperature in Eq.(\ref{eq:TDI}) can be performed by combining the CEH with atomistic Monte Carlo (AMC) technique.

The CEH is constructed by mapping a large database of Density Functional Theory (DFT) energies of not only randomized solution structures but also for intermetallic ordered structures with the different compositions \cite{ducnm2021,muzyk2011,damian2020,andrew2023}.  Within the present study for W--Ta--V alloys, the set of ECIs energies obtained from our recent study \cite{andrew2023} have been employed as it takes into consideration not only two and three but also four-body cluster interaction which is crucial in discovering the important role of the CSRO V-Ta CSRO parameter. In this work, semi-canonical exchange AMC simulations were performed using Alloy Theoretic Automated toolkit (ATAT) \cite{vdw2002a,vdw2002b} package for the BCC $20 \times 20 \times 20$ super-cell with 16000 atoms. For each composition of the ternary W--Ta--V alloys, the simulation computes the enthalpy of mixing at each temperature step for reaching a thermodynamic equilibrium  configuration. The start temperature for each simulation begins at 2000 K and is reduced to 10 K with temperature steps of 10 K. The start temperature is chosen such that a disordered configuration is near-guaranteed after 2000 MC steps per atom in the thermalization and accumulation stages. The free energy of an alloy at each temperature is calculated from the enthalpy of mixing and the configuration entropy determined by Eq.(\ref{eq:TDI}). The error bars of CSRO parameters can be determined by Eq.(\ref{eq:sromatrix}) using AMC simulations of average cluster functions calculated from thermodynamic equilibrium configurations generated for each temperature and alloy composition.
	
\section*{Acknowledgements}
\noindent Research presented in this article was supported by the Laboratory Directed Research and Development (LDRD) program of Los Alamos National Laboratory primarily under project number 20220597ECR. MAT acknowledges support from the LDRD program 20200689PRD2. This work was performed, in part, at the Center for Integrated Nanotechnologies, an Office of Science User Facility operated for the U.S. Department of Energy (DOE) Office of Science. Los Alamos National Laboratory, an affirmative action equal opportunity employer, is managed by Triad National Security, LLC for the U.S. Department of Energy NNSA, under contract 89233218CNA000001. DNM, JSW and DS work has been carried out within the framework of the EUROfusion Consortium, funded by the European Union via the Euratom Research and Training Programme (Grant Agreement No 101052200 - EUROfusion). The work at UKAEA was partially supported by the Broader Approach Phase II agreement under the PA of IFERC2-T2PA02. Views and opinions expressed are however those of the author(s) only and do not necessarily reflect those of the European Union or the European Commission. Neither the European Union nor the European Commission can be held responsible for them. DNM also acknowledges funding by the EPSRC Energy Programme [grant number EP/W006839/1]. The work at WUT has been carried out as a part of an international project co-financed from the funds of the program of the Polish Minister of Science and Higher Education through the PMW program in 2023. DNM and JSW would like to thank the support from high-performing computing facility MARCONI (Bologna, Italy) provided by EUROfusion. All the authors are grateful for funding provided by the Austrian Research Promotion Agency (FFG) in the project 3DnanoAnalytics (FFG-No 858040).
	
\section*{CRediT author statement}
\noindent - Conceptualization and Visualization: \\
\noindent - Methodology, Investigation, Data Curation, Formal Analysis:  \\
\noindent - Resources:  \\
\noindent - Supervision and Funding Acquisition: \\
\noindent - Writing Original Draft: . \\
\noindent - Writing Review and Editing: All authors. \\
\noindent - Project administration: .
	
\section*{Data availability}
\noindent The raw data required to reproduce these findings are available in the Mendeley Data permanent repository: \href{https://doi.org/10.17632/y326fsd8ww.1}{10.17632/y326fsd8ww.1}.
	
\bibliographystyle{naturemag}
\bibliography{bibdata.bib}
	
\end{document}